# Adaptive prescribed-time disturbance observer using nonsingular terminal sliding mode control: Extended Kalman filter and particle swarm optimization

Amin Vahidi-Moghaddam[1], Arman Rajaei[2], Moosa Ayati[3*], Ramin Vatankhah[2], Mohammad Reza Hairi-Yazdi[3]

[1] Department of Mechanical Engineering, Michigan State University, East Lansing, Michigan, USA
[2] School of Mechanical Engineering, Shiraz University, Shiraz, Fars, Iran
[3] Department of Mechanical Engineering, University of Tehran, Tehran, Tehran, Iran
* m.ayati@ut.ac.ir

**Abstract:** In this paper, adaptive prescribed finite time stabilization of uncertain single-input and single-output nonlinear systems is considered in the presence of unknown states, unknown parameters, external load disturbance, and non-symmetric input saturation. A prescribed finite time disturbance observer is designed to approximate the unmeasured external disturbance. Also, a nonsingular prescribed finite time terminal sliding mode control is proposed for the closed-loop control of the system with the non-symmetric input saturation. Extended Kalman filter algorithm is employed for the real-time estimations of the states and unknown parameters of the system. Moreover, particle swarm optimization algorithm is used to obtain the design parameters of the proposed disturbance observer and controller. To show the performance of designed control scheme, the proposed approach is employed to guarantee prescribed finite time stabilization of nonlinear vibration of a nonlocal strain gradient nanobeam. The Galerkin projection method is used to reduce the non-dimensional form of the governing nonlinear partial differential equation of Euler–Bernoulli nanobeam to the ordinary differential equation. Finally, numerical simulations are performed to illustrate the effectiveness and performance of the developed adaptive control scheme for the vibration control of nanobeam in comparison to the conventional sliding mode control.

## 1. Introduction

Sliding mode control (SMC) is one of the most popular controllers due to its unique characteristics for decreasing the tracking error of nonlinear systems with uncertainty and disturbance. Recently, variants of SMC, such as neural SMC [1], event-triggered SMC [2], second order SMC [3], discrete-time SMC [4], adaptive SMC [5], and integral SMC [6] have been successfully used for robust stabilization of different systems. However, the basic SMC may not ensure the prescribed finite time convergence of the entire closed-loop error signals to zero. Thus, terminal sliding mode control (TSMC) has been developed to improve the conventional SMC and achieve the prescribed finite time stability. For many control problems, variants of TSMC, such as nonsingular TSMC [7], fast TSMC [8], and neural TSMC [9] have guaranteed the finite time stabilization of the system. However, there are some shortcomings for the existing research results: 1) The states and parameters of the system are assumed to be measurable, but they may not be available for measurement in the most practical application scenarios. 2) The upper bound of external disturbance is directly used to design TSMC, but it causes excessively conservative controllers and unnecessary large control efforts. 3) The control input is allowed to have large amounts, but it should be in a specific range for the practical applications. 4) To develop the TSMC and disturbance observer, all of the design parameters should be in their allowable limits in order to accomplish a good performance, but it is a time-consuming procedure to choose the best range for each design parameter in which the TSMC stabilizes the system in the minimum possible time. To obviate these issues and improve the performance of the TSMC, this controller needs to be investigated by a new technique.

Disturbance rejection is used to improve the performance of control scheme for stabilization of the systems in the presence of external disturbance [10-12]. However, a disturbance observer is employed to explore the information about the characteristic of external disturbance and estimate it; consequently, it improves the performance of controllers. Variants of disturbance observer, such as fractional-order disturbance observer [13], self-learning disturbance observer [14], disturbance observer-based integral backstepping control [15], disturbance observer-based backstepping SMC [16], and disturbance observer-based global SMC [17] have used to estimate the disturbance and improve the controller. However, the existing research results does not guarantee the prescribed finite time convergence of disturbance approximation error to zero.

Extended Kalman filter (EKF) algorithm is one of the useful approaches to estimate the states of nonlinear systems [18,19]. To apply this method, the system must be linearized in each time step, and last observations must be used in a recursive formula to estimate the actual amount of states while the influence of noise is reduced. Moreover, this method is helpful in the estimation of parameters of system when the system is nonlinear to them. In this case, using the EKF algorithm, the augmented state vector must be defined by states and unknown parameters of the system to construct an augmented nonlinear system. The EKF has been used for state estimation of non-classical microcantilevers in [20]. Invariant extended Kalman filter on matrix Lie groups has been developed in [21]. In this paper, the EKF algorithm is utilized to estimate the states and unknown parameters of system.

Particle swarm optimization (PSO) algorithm is employed to obtain an optimal solution for real word optimization problems. Recently, the PSO algorithm has been successfully used to improve the performance of control schemes. Variants of PSO algorithm, such as multi-objective control-based PSO algorithm [22]. system identification-based PSO algorithm [23], coverage control-based PSO algorithm [24], and fuzzy SMC-based PSO algorithm [25] have been successfully developed for control of various systems. In this paper, the PSO algorithm is employed to achieve the design parameters of the proposed disturbance observer-based TSMC.

Recently, micro/nano-scale beams play significant roles in micro/nano-electromechanical systems (MEMS/NEMS) such as nanocomposites [26], micro-actuators [27], micro-resonators [28], and so on [29,30]. Nonlinear vibration is one of the most important problems in micro/nano-scale beams. A beam with axially immovable ends has the nonlinear static or vibration behaviour for the large transverse loads, since the large deflection causes the axial tension while the strains have to remain small. Consequently, many studies have been conducted to study the vibration behaviour of micro/nanobeams [31-35]. Furthermore, the analysis of vibration control is used to prevent the micro/nanobeams from damage and improve the performance and resolution of beams [36-39]. However, the measured dimensions of nanobeams may not be exact; therefore, the estimation analysis and robust controllers are useful to enhance the accuracy. Moreover, prescribed finite time stabilization and disturbance observer improve the results for vibration control of micro/nanobeams.

In this paper, an adaptive nonsingular prescribed finite time disturbance observer-based TSMC is developed to stabilize the uncertain single-input and single-output (SISO) nonlinear systems. The EKF is used for a state/parameter augmented system to estimate unmeasurable states and unknown parameters from noisy output. A disturbance observer is designed to approximate an unmeasured external disturbance in prescribed finite time. Using the output of disturbance observer, the nonsingular TSMC is developed for stabilization of uncertain nonlinear systems. The prescribed finite time stability of the closed-loop system is proved via Lyapunov stability theorem. To satisfy control input constraint, the controller formula, stability, and convergence of the closed-loop system are reformulated for an unknown non-symmetric input saturation. The PSO algorithm is employed to obtain the optimal design parameters of the developed disturbance observer-based TSMC by considering the stabilization time as the cost function. The designed control scheme is employed to suppress nonlinear vibration of Euler–Bernoulli nonlocal strain gradient nanobeams. Finally, numerical results are presented to confirm the effectiveness of designed control scheme for the vibration control of nanobeam.

The present paper is organized as follows. Problem statement and preliminaries are presented in Section 2. The design of nonsingular prescribed finite time disturbance observer-based TSMC is presented in Section 3. In Section 4, the EKF algorithm is presented for the real-time estimations of states and unknown parameters of the system. The PSO algorithm is employed to obtain the optimal design parameters of the proposed disturbance observer-based TSMC in Section 5. The modelling of Euler-Bernoulli nanobeam under a centralized force in the middle of the beam is presented in Section 6. In Section 7, the numerical simulations are presented, followed by the conclusion are drawn in Section 8.

## 2. Problem Formulation and Preliminaries

In this section, the problem is formulated for the adaptive stabilization of SISO uncertain nonlinear systems in prescribed finite time. Moreover, some basic preliminaries are provided for the prescribed finite time convergence.

### 2.1. Problem Statement

Consider the following state space dynamics for the SISO uncertain nonlinear systems exposed by the external disturbance as

$$\begin{cases} \dot{x}_i = x_{i+1}, & i = 1,2,\ldots,n-1 \\ \dot{x}_n = f(x) + g(x)u + d \\ y = x_1 \end{cases} \quad (1)$$

where $x \in \mathbb{R}^n$ and $f(x) \in \mathbb{R}$ stand for the states and drift dynamics of the system, $u \in \mathbb{R}$ and $g(x) \in \mathbb{R}$ denote the control input and input dynamics of the system, $d$ is the external disturbance, and $y \in \mathbb{R}$ denotes the output of the system.

**Assumption 1.** The functions $f(x)$ and $g(x)$ are locally Lipschitz, and $f(0) = 0$.

**Assumption 2.** The structure of the system dynamics $f$ is known; however, it encompasses unknown parameters.

**Assumption 3.** The states of the system are not available for measurement.

**Assumption 4.** The unknown external disturbance $d$ is bounded, i.e. $|d| < \beta_0$, where $\beta_0$ is a positive parameter.

**Remark 1**. Assumption 1 is a standard assumption to guarantee the existence and uniqueness of the solution for (1). Assumption 2 and Assumption 3 help us to achieve a suitable control scheme for practical applications. To observe the disturbance, we consider a bounded unknown external disturbance.

**Lemma 1 [40].** If $V(t)$ is a continuous positive definite function that satisfies

$$\dot{V}(t) + \vartheta V(t) + \xi V^\gamma(t) \le 0, \ \forall t > t_0 \quad (2)$$

thus, $V(t)$ converges to the equilibrium point in prescribed finite time

$$t_s \le t_0 + \frac{1}{\vartheta(1-\gamma)} \ln \frac{\vartheta V^{1-\gamma}(t_0) + \xi}{\xi} \quad (3)$$

where $\vartheta, \xi$, and $0 < \gamma < 1$ are positive parameters. □

**Lemma 2 [7,40].** Let $V(t)$ as a continuous function. Thus,

$$\frac{d^{(n-j)}}{dt^{(n-j)}} V_j^{\gamma_j}, \quad j = 1,\ldots,n \quad (4)$$

is bounded for $V_j = 0$ and $0 < \gamma_j < 1$ if

$$\gamma_j > \frac{n-j}{n-j+1} \quad (5)$$

Therefore, $\gamma_j$ must satisfy both $0 < \gamma_j < 1$ and (5). □

Let $y_d$ and $\hat{d}$ as the desired output and the estimated external disturbance for the system (1). The problem of interest in this paper is formulated as

**Problem 1 (Adaptive disturbance observer using TSMC).** Consider the system (1), design a nonsingular disturbance observer-based control such that the following properties are satisfied:

**Property 1.** The disturbance approximation error $\tilde{d} = \hat{d} - d$ converges to zero in prescribed finite time.

**Property 2.** The tracking error $e = y - y_d$ converges to zero in prescribed finite time.

**Remark 2.** In Problem 1, Property 1 and Property 2 mean that the adaptive nonsingular disturbance observer-based TSMC stabilizes the uncertain nonlinear system (1) in prescribed finite time.

## 3. Controller Design

In this section, a nonsingular prescribed finite time disturbance observer-based TSMC is designed for the system (1) with non-symmetric input saturation.

### 3.1. Prescribed Finite Time Disturbance Observer using TSMC

The disturbance observer with the prescribed finite time convergence property is designed as follows [41]
$$\hat{d} = -ks - \beta_0 \, sgn(s) - \varepsilon s^{p_0/q_0} - |f(x)| \, sgn(s) - f(x) \quad (6)$$
where $\hat{d}$ is the estimated external disturbance, $\beta_0$ is the bound of external disturbance according to Assumption 4, $k$ and $\varepsilon$ are positive design parameters, $p_0$ and $q_0$ are odd positive integers such that $p_0 < q_0$, and $s$ is an auxiliary variable as
$$s = z - x_n \quad (7)$$
where $z$ is obtained as
$$\dot{z} = -ks - \beta_0 \, sgn(s) - \varepsilon s^{p_0/q_0} - |f(x)| \, sgn(s) + g(x)u \quad (8)$$

Using (1) and (6)-(8), the disturbance approximation error is
$$\tilde{d} = \hat{d} - d = \dot{z} - \dot{x}_n = \dot{s} \quad (9)$$

**Theorem 1.** Consider the system (1). The disturbance observer (6)-(8) makes the prescribed finite time convergence of the disturbance approximation error (9) to zero.

**Proof.** Let $V_0 = \frac{1}{2}s^2$ as the Lyapunov function candidate. The time derivative of the Lyapunov function candidate is "(see (10))"

Therefore, according to Lemma 1 and (9), the disturbance approximation error $\tilde{d}$ converges to zero in prescribed finite time, and the proof is completed. □

**Remark 3.** Using (10), $s$ satisfies Lemma 1; therefore, it converges to zero in prescribed finite time and remains at zero. Consequently, $\dot{s}$ converges to zero in prescribed finite time. Now, using (9), $\tilde{d}$ converges to zero in prescribed finite time. Thus, using Theorem 1, the disturbance observer (6)-(8) guarantees the prescribed finite time convergence of the disturbance approximation error to zero for the uncertain nonlinear system (1).

To design the nonsingular prescribed finite time TSMC using the output of the disturbance observer, the following variables should be defined for the uncertain nonlinear system (1).
$$s_1 = y - y_d = x_1 - y_d \quad (11)$$
$$s_1^{(n)} = y^{(n)} - y_d^{(n)} = \dot{x}_n - y_d^{(n)} \quad (12)$$

Then, with the recursive procedure, the following equations is obtained as
$$s_2 = \dot{s}_1 + \alpha_1 s_1 + \beta_1 s_1^{p_1/q_1}$$
$$s_3 = \dot{s}_2 + \alpha_2 s_2 + \beta_2 s_2^{p_2/q_2} \quad (13)$$
$$\vdots$$
$$s_n = \dot{s}_{n-1} + \alpha_{n-1} s_{n-1} + \beta_{n-1} s_{n-1}^{p_{n-1}/q_{n-1}} + s$$

where $\alpha_j$ and $\beta_j$ are positive design parameters, and $p_j$ and $q_j$ are odd positive integers ($p_j < q_j$). Therefore, using (1), (9), and (11)-(13), we have "(see (14))"

**Theorem 2.** Consider the system (1) with $g(x) \neq 0$. The nonsingular TSMC law
$$u = -\frac{1}{g(x)}\left(f(x) - y_d^{(n)} + \sum_{j=1}^{n-1} \alpha_j s_j^{(n-j)} + \sum_{j=1}^{n-1} \beta_j \frac{d^{(n-j)}}{dt^{(n-j)}} s_j^{p_j/q_j} + \hat{d} + \delta s_n + \mu s_n^{p_n/q_n}\right) \quad (15)$$
guarantees the prescribed finite time convergence of the states of the system (1) to zero, where $\delta$ and $\mu$ are positive design parameters.

**Proof.** The control law (15) is designed to satisfy the conditions of Lemma 1; therefore, using (9), substituting (15) into (14) yields
$$\dot{s}_n = -\delta s_n - \mu s_n^{p_n/q_n} \quad (16)$$
Let $V = \frac{1}{2} s_n^2$ as the Lyapunov function candidate. The time derivative of the Lyapunov function candidate is given as
$$\dot{V} \leq -\delta s_n^2 - \mu s_n^{(p_n + q_n)/q_n}$$
$$\leq -2\delta V - \mu 2^{(p_n+q_n)/2q_n} V^{(p_n+q_n)/2q_n} \quad (17)$$
Thus, according to Lemma 1, the proof is completed. □

**Remark 4.** Using the output of the disturbance observer (6), the nonsingular TSMC (15) stabilizes the uncertain nonlinear system (1) in prescribed finite time. Based on Lemma 2, the term $p_j/q_j$ in (15) must satisfy (5) to avoid the singularity problem in the control law for $s_j = 0$. It is worth nothing that the prescribed finite time convergence of $s_n$ is guaranteed using (17), then using (13) and Theorem 1, it is guaranteed that $s_{n-1}$ satisfies Lemma 1 and converges to zero in prescribed finite time. Therefore, we have a similar procedure for $s_{n-2}, \ldots, s_2, s_1$; consequently, the prescribed finite time stabilization of system (1) is guaranteed.

### 3.2. Prescribed Finite Time Disturbance Observer using TSMC with Non-symmetric Input Saturation

In this section, an unknown non-symmetric input saturation is considered for the system (1). Modifications are performed on the formulation, and the stability proof is developed for the proposed controller in the presence of non-symmetric input saturation. Assume the control input as
$$u = \begin{cases} u_{\max} & u_c > u_{\max} \\ u_c & u_{\min} \leq u_c \leq u_{\max} \\ u_{\min} & u_c < u_{\min} \end{cases} \quad (18)$$
where $u_{\min}$ and $u_{\max}$ are the lower and upper bounds of the non-symmetric input saturation. The designed control input $u_c$ is designed as
$$u_c = g(x)(g^2(x) + \tau)^{-1} v_r \quad (19)$$
where $\tau$ is a positive design parameter, and $v_r$ will be obtained later.

Substituting (18) and (19) into (1) yields
$$\dot{x}_n = f(x) + g(x)(u_c + \Delta u) + d$$
$$= f(x) + v_r + d + g(x)\Delta u - \tau(g^2(x) + \tau)^{-1} v_r \quad (20)$$
$$= f(x) + v_r + D$$
where $\Delta u = u - u_c$ cannot be calculated since $u_{\min}$ and $u_{\max}$ are unknown parameters. Hence, the compound disturbance $D$ is defined to handle this problem when the input is saturated.

**Assumption 5.** The defined compound disturbance $D$ is bounded, i.e. $|D| < \beta_0$.

$$\begin{aligned}\dot{V}_0 = s\dot{s} &= s\left(-ks - \beta_0\, sgn(s) - \varepsilon s^{p_0/q_0} - |f(x)|\, sgn(s) - f(x) - d\right) \\ &\leq -ks^2 - \beta_0\, s\, sgn(s) - \varepsilon s^{(p_0+q_0)/q_0} + |s||d| \\ &\leq -2kV_0 - 2^{(p_0+q_0)/2q_0}\varepsilon\, V_0^{(p_0+q_0)/2q_0}\end{aligned} \qquad (10)$$

$$\begin{aligned}\dot{s}_n &= s_1^{(n)} + \sum_{j=1}^{n-1}\alpha_j s_j^{(n-j)} + \sum_{j=1}^{n-1}\beta_j \frac{d^{(n-j)}}{dt^{(n-j)}} s_j^{p_j/q_j} + \dot{s} \\ &= \dot{x}_n - y_d^{(n)} + \sum_{j=1}^{n-1}\alpha_j s_j^{(n-j)} + \sum_{j=1}^{n-1}\beta_j \frac{d^{(n-j)}}{dt^{(n-j)}} s_j^{p_j/q_j} + \dot{s}\end{aligned} \qquad (14)$$

The disturbance observer with the prescribed finite time convergence property for the unknown compound disturbance $D$ is designed as
$$\widehat{D} = -ks - \beta_0\, sgn(s) - \varepsilon s^{p_0/q_0} - |f(x)|\, sgn(s) - f(x) \qquad (21)$$
where $s$ is
$$s = z - x_n \qquad (22)$$
and $z$ is
$$\dot{z} = -ks - \beta_0\, sgn(s) - \varepsilon s^{p_0/q_0} - |f(x)|\, sgn(s) + v_r \qquad (23)$$

**Theorem 3.** Consider the system (1). The nonsingular TSMC law
$$v_r = -f(x) + y_d^{(n)} - \sum_{j=1}^{n-1}\alpha_j s_j^{(n-j)} - \sum_{j=1}^{n-1}\beta_j \frac{d^{(n-j)}}{dt^{(n-j)}} s_j^{p_j/q_j} - \widehat{D} - \delta s_n - \mu s_n^{p_n/q_n} \qquad (24)$$
guarantees the prescribed finite time convergence of the states of the system (1) to zero in the presence of non-symmetric input saturation.

**Proof.** Like the previous subsection, $\widetilde{D} = \dot{s}$ and $\dot{s}_n = -\delta s_n - \mu s_n^{p_n/q_n}$. Considering $V = \frac{1}{2}s_n^2$ as the Lyapunov function candidate, its time derivative is
$$\begin{aligned}\dot{V} &\leq -\delta s_n^2 - \mu s_n^{(p_n+q_n)/q_n} \\ &\leq -2\delta V - \mu 2^{(p_n+q_n)/2q_n} V^{(p_n+q_n)/2q_n}\end{aligned} \qquad (25)$$
Thus, $V$ satisfies Lemma 1, and the proof is completed. □

**Remark 5.** For the unknown non-symmetric input saturation (18), the states of the closed-loop system converge to zero in prescribed finite time using the nonsingular control law (24). The term $p_j/q_j$ must satisfy (5) to avoid the singularity problem in the control law (24) for $s_j = 0$. The design parameter $\beta_0$ should be chosen as a large positive constant to guarantee the design requirement of the proposed control scheme.

## 4. Extended Kalman Filter Algorithm

In this section, the EKF algorithm is utilized to estimate the states and unknown parameters of the system (1).

In the first step, the system (1) should be discretized as
$$\begin{aligned}x_k &= f_{cl}(x_{k-1}) + w_{k-1} \\ y_k &= h(x_{k-1}) + v_{k-1}\end{aligned} \qquad (26)$$
with
$$\dot{x} = \frac{x_k - x_{k-1}}{T_s} \qquad (27)$$
where $T_s$ is the sample time. $w_{k-1}$ and $v_{k-1}$ are the zero mean Gaussian white noise processes. $w_{k-1}$ is the process noise and relates to the quality of the model, and $v_{k-1}$ is the measurement noise and relates to the quality of the measurement system.

Following equations show the predictive phase (28) and the correction phase (29) of the state estimation vector $\hat{x}$ and the covariance matrix $P$.
$$\begin{aligned}\hat{x}_{k|k-1} &= f_{cl}(\hat{x}_{k-1|k-1}) \\ P_{k|k-1} &= F_{k-1}\, P_{k-1|k-1}\, F_{k-1}^T + Q_{k-1}\end{aligned} \qquad (28)$$
$$\begin{aligned}B_k &= y_k - h(\hat{x}_{k|k-1}) \\ S_k &= H_k\, P_{k|k-1}\, H_k^T + R_k \\ K_k &= P_{k|k-1}\, H_k^T\, S_k^{-1} \\ \hat{x}_{k|k} &= \hat{x}_{k|k-1} + K_k B_k \\ P_{k|k} &= P_{k|k-1} - K_k\, S_k\, K_k^T\end{aligned} \qquad (29)$$
where $Q$ and $R$ are the covariance matrices of the process and measurement noises, respectively. $K_k$ is the Kalman gain, and $F_k$ and $H_k$ are the Jacobin matrices of $f_{cl}(x)$ and $h(x)$ as follows
$$\begin{aligned}[F_k]_{i,j} &= \left.\frac{\partial f_{cl\,i}(x)}{\partial x_j}\right|_{x=\hat{x}_{k|k}} \\ [H_k]_{i,j} &= \left.\frac{\partial h_i(x)}{\partial x_j}\right|_{x=\hat{x}_{k|k-1}}\end{aligned} \qquad (30)$$

In this work, using the EKF, the augmented state vector is defined by states and unknown parameters of the system (1) to construct an augmented nonlinear system. The EKF is utilized for simultaneous state and parameter estimation of the system (1). The separation principle holds for affine nonlinear systems using an exponentially stable observer. As an exponentially stable observer, the extended Kalman filter algorithm presents an estimation error which does not depend on control input for affine nonlinear systems [42,43]; therefore, we can use the separation principle only for this case. Based on the separation principle, the estimated states are employed to design the disturbance observer and the controller such that last observations are used in a recursive formula to estimate the actual amount of states. Thus, the system (1) with unknown parameters is stabilized without requiring the states of the system. The details are presented in Subsection 5.3.

As is shown in the block diagram of Fig. 1, the estimated parameters, the estimated states, and the estimated disturbance are employed in the adaptive nonsingular prescribed finite time TSMC for the uncertain nonlinear system (1) with unknown non-symmetric input saturation.

## 5. Particle Swarm Optimization Algorithm

In this section, the PSO algorithm is employed to obtain the optimal design parameters of the developed disturbance observer-based TSMC by considering the stabilization time as the cost function.

Based on the simulation of birds flocking, the PSO is developed to optimize a certain cost function. Each bird

(particle) in the swarm demonstrates a candidate solution to the optimization problem, and its position is presented by a point in the search space. For an optimization problem with $N$ variables, a population with the size of particles is created, and their positions are expressed as

$$X = [X_1, X_2, \ldots, X_N] \tag{31}$$

In the PSO, each particle flies through the search space and regulates its position using their personal and global experiences. Each particle records its best personal position (personal experience) and best position in the group (global experience) so far. Each particle modifies its position using the concept of velocity as follows

$$V_{k+1} = W V_k + R_1 C_1 (P_k - X_k) + R_2 C_2 (G_k - X_k) \tag{32}$$

where $P_k$ and $G_k$ stand for the best personal and global experiences, respectively. $W$ indicates the inertia weight which illustrates the tendency of an object for balance. $C_1$ and $C_2$ are two positive parameters and represent the personal and global learning coefficients, respectively. Indeed, these parameters determine the significance of the particle's best personal and global experiences, respectively. In addition, $R_1$ and $R_2$ denote uniform random numbers, and are bounded within $[0, 1]$.

Using (32), the new position of each particle is obtained as

$$X_{k+1} = X_k + V_{k+1} \tag{33}$$

The next iteration takes place after moving all particles. Consequently, the swarm is moved close to an optimum of the fitness function which is evaluated based on the cost function of optimization problem. The PSO is performed until a predefined number of generations is achieved.

It is worth noting that to initialize the PSO, the positions of particles are allocated to random locations in the search space. Also, their velocities are chosen in the range $[V_{min}, V_{max}]$. Moreover, the velocity of each particle is limited by $V_{max}$ at each time step.

In this work, the design parameters of the proposed disturbance observer-based TSMC are considered as the particles. Therefore, $N$ shows the number of design parameters. Moreover, the stabilization time is considered as the cost function of this optimization problem. Random location is considered for the position (value) of each design parameter to initialize the PSO. Using (32) and (33), the PSO updates the value of each design parameter and evaluates the optimum of the fitness function based on the stabilization time. The procedure is summarized in Table 1.

Table 1 Particle swarm optimization algorithm

| | |
|---|---|
| 1- | Initialize a population array of particles. |
| 2- | Loop |
| 3- | Evaluate the fitness function for each particle. |
| 4- | Modify the particle's fitness. |
| 5- | Identify the best particle in the neighbourhood. |
| 6- | Calculate the velocities of particles. |
| 7- | Update the positions of particles. |
| 8- | Exit loop if the number of generations is met. |
| 9- | End |

## 6. Nonlocal Strain Gradient Nanobeam

Consider a simply-supported Euler-Bernoulli nanobeam with immovable ends of length L, width b, and thickness h under a centralized force in the middle of the beam, as shown in Fig. 2. Based on the nonlocal strain gradient theory, the governing partial nonlinear differential equation of the Euler–Bernoulli nanobeam is obtained by employing Hamilton's principle as [28] "(see (34))" where $w(x,t)$ is the lateral deflection, $x$ and $t$ denote the independent spatial and time variables, $I_A$ is the axial inertia coefficient of the nanobeam, $ea$ indicates the nonlocal parameter to consider the significance of nonlocal elastic stress field, and $l_m$ is the strain gradient length scale parameter. $D_{xx}$, $A_{xx}$, and $I_A$ are obtained for a homogenous beam as

$$D_{xx} = EI, \quad A_{xx} = EA, \quad I_A = \rho A \tag{35}$$

$P_0(x,t)$ is a centralized force in the middle of the nanobeam as

$$P_0(x,t) = P_0(t)\delta\left(x - \frac{L}{2}\right) \tag{36}$$

(34) is rewritten in the non-dimensional form to reduce the numerical errors during the simulations. Thus, the following standard dimensionless quantities are brought up for a beam with a rectangular cross section as

$$\bar{x} = \frac{x}{L}, \; \bar{w} = \frac{w}{k_0}, \; \bar{z} = \frac{z}{h}, \; \bar{t} = t\sqrt{\frac{EI}{\rho A L^4}}, \; \alpha = \frac{ea}{L}, \; \beta = \frac{l_m}{L},$$

$$\bar{\delta}\left(\bar{x} - \frac{1}{2}\right) = L\delta\left(x - \frac{L}{2}\right) \tag{37}$$

where $I$ is the inertia moment of the cross-section, $k_0 = \sqrt{I/A}$ is the radius gyration, and $E$, $\rho$, and $A$ represent the Young's modulus, mass density, and cross-sectional area, respectively. Substituting (37) into (34), the dimensionless form of the governing equation is expressed as "(see (38))"

$$D_{xx} l_m^2 \frac{\partial^6 w}{\partial x^6} - D_{xx} \frac{\partial^4 w}{\partial x^4} + \left[\frac{A_{xx}}{2L} \int_0^L (\frac{\partial w}{\partial x})^2 dx - \frac{A_{xx} l_m^2}{L} \int_0^L (\frac{\partial w}{\partial x} \frac{\partial^3 w}{\partial x^3} + (\frac{\partial^2 w}{\partial x^2})^2) dx\right] \times$$
$$\left[\frac{\partial^2 w}{\partial x^2} - (ea)^2 \frac{\partial^4 w}{\partial x^4}\right] + I_A \frac{\partial^2}{\partial t^2}\left[(ea)^2 \frac{\partial^2 w}{\partial x^2} - w\right] = (ea)^2 \frac{\partial^2 P_0}{\partial x^2} - P_0 \tag{34}$$

$$\beta^2 \bar{D}_{xx} \frac{\partial^6 \bar{w}}{\partial \bar{x}^6} - \bar{D}_{xx} \frac{\partial^4 \bar{w}}{\partial \bar{x}^4} + \left[\frac{\bar{A}_{xx}}{2} \int_0^1 (\frac{\partial \bar{w}}{\partial \bar{x}})^2 d\bar{x} - \beta^2 \bar{A}_{xx} \int_0^1 (\frac{\partial \bar{w}}{\partial \bar{x}} \frac{\partial^3 \bar{w}}{\partial \bar{x}^3} + (\frac{\partial^2 \bar{w}}{\partial \bar{x}^2})^2) d\bar{x}\right] \frac{\partial^2 \bar{w}}{\partial \bar{x}^2}$$
$$- \left[\frac{\alpha^2 \bar{A}_{xx}}{2} \int_0^1 (\frac{\partial \bar{w}}{\partial \bar{x}})^2 d\bar{x} - \alpha^2 \beta^2 \bar{A}_{xx} \int_0^1 (\frac{\partial \bar{w}}{\partial \bar{x}} \frac{\partial^3 \bar{w}}{\partial \bar{x}^3} + (\frac{\partial^2 \bar{w}}{\partial \bar{x}^2})^2) d\bar{x}\right] \frac{\partial^4 \bar{w}}{\partial \bar{x}^4} + \alpha^2 \bar{I}_A \frac{\partial^4 \bar{w}}{\partial \bar{x}^2 \partial \bar{t}^2} \tag{38}$$
$$- \bar{I}_A \frac{\partial^2 \bar{w}}{\partial \bar{t}^2} = \alpha^2 \lambda \frac{\partial^2 \bar{P}_0}{\partial \bar{x}^2} - \lambda \bar{P}_0$$

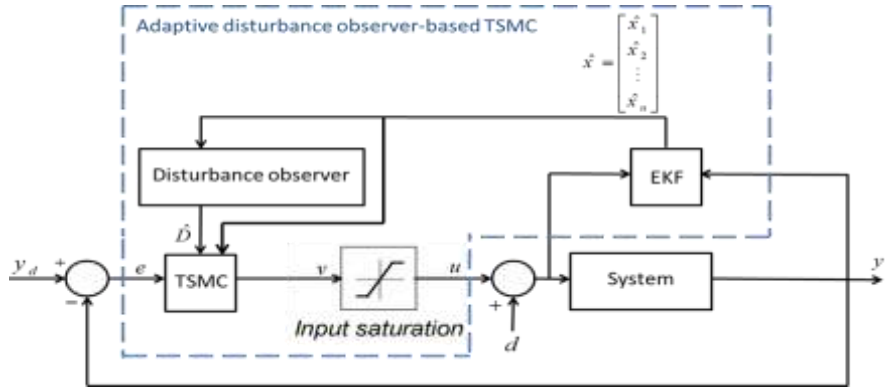

**Fig. 1.** Closed-loop system with the adaptive prescribed finite time disturbance observer using nonsingular TSMC.

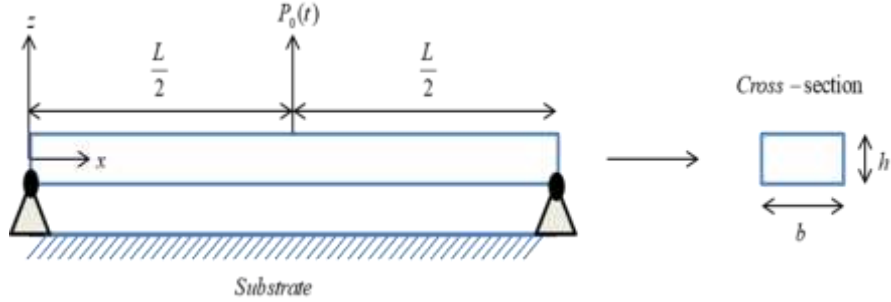

**Fig. 2.** Simply-supported nanobeam with immovable ends under the centralized force in the middle of beam ($P_0$).

where $\lambda = \frac{12\, L^3}{E\, A\, h^2\, r}$ and $\bar{P}_0(\bar{x}, \bar{t}) = P_0(\bar{t})\, \bar{\delta}(\bar{x} - \frac{1}{2})$. For the homogenous nanobeam, $\bar{D}_{xx}$, $\bar{A}_{xx}$, and $\bar{I}_A$ are obtained as follows

$$\bar{D}_{xx} = 1, \bar{A}_{xx} = 1, \bar{I}_A = 1 \tag{39}$$

Now, Galerkin projection method is employed to reduce the above partial differential equation to the ordinary differential equation. To do so, the temporal and spatial terms of $\bar{w}(\bar{x}, \bar{t})$ are decomposed, and the lateral deflection of beam is expanded as follows

$$\bar{w}(\bar{x}, \bar{t}) = Q(\bar{t})\phi(\bar{x}) \tag{40}$$

where $Q(\bar{t})$ is the unknown temporal part of lateral deflection of the nanobeam, and $\phi(\bar{x})$ is its spatial part which must satisfy the kinematic boundary conditions of the nanobeam. For the simply-supported nanobeam, $\phi(\bar{x})$ is considered as

$$\phi(\bar{x}) = \sin(\pi \bar{x}) \tag{41}$$

The following equations are considered for the centralized force in the middle of the beam as

$$\int_{a-\varepsilon}^{a+\varepsilon} f(x)\, \delta(x-a)dx = f(a), \quad \int_{a-\varepsilon}^{a+\varepsilon} f(x)\, \delta^{(n)}(x-a)\, dx = -\int_{a-\varepsilon}^{a+\varepsilon} \frac{\partial f}{\partial x} \delta^{(n-1)}(x-a)dx \tag{42}$$

Substituting (41) and (42) into (38), multiplying both sides by $\phi(\bar{x})$, and integrating the obtained equation over the beam length ($\bar{x} = 0$ to $\bar{x} = 1$), the following dimensionless ordinary differential equation is achieved as

$$\ddot{Q}(\bar{t}) + K_1 Q(\bar{t}) + K_2 Q^3(\bar{t}) = -g P_0(\bar{t}) \tag{43}$$

where the coefficients $K_1, K_2$, and $g$ are "(see (44-46))"

$$K_1 = \frac{\beta^2 \bar{D}_{xx} \int_0^1 \phi^{(6)} \phi\, d\bar{x} - \bar{D}_{xx} \int_0^1 \phi^{(4)} \phi\, d\bar{x}}{\alpha^2 \bar{I}_A \int_0^1 \phi'' \phi\, d\bar{x} - \bar{I}_A \int_0^1 (\phi)^2\, d\bar{x}} \tag{44}$$

$$g = \frac{\lambda\, (\alpha^2 \pi^2 + 1)}{\alpha^2 \bar{I}_A \int_0^1 \phi'' \phi\, d\bar{x} - \bar{I}_A \int_0^1 (\phi)^2\, d\bar{x}} \tag{46}$$

Let $x_1 = Q(\bar{t})$, $x_2 = \dot{x}_1 = \dot{Q}(\bar{t})$, and $u = P_0(\bar{t})$. (43) is rewritten in the state space form as

$$\begin{cases} \dot{x}_1 = x_2 \\ \dot{x}_2 = -K_1 x_1 - K_2 x_1^3 - gu + d(\bar{t}) \end{cases} \tag{47}$$

**Assumption 6.** The coefficients $K_2$ and $g$ are known, but the coefficient $K_1$ is an unknown parameter.

**Assumption 7.** $y_d = 0$ is the desired system output for the nanobeam.

The EKF algorithm is employed to estimate the states of the system ($x_1$ and $x_2$), and the unknown parameter $K_1$.

## 7. Simulation Study

In this section, the simulation results are presented for the state space system (47), where $K_1 = 97.4$, $K_2 = -19.97$, and $g = -1.09$. It is worth noting that the design parameters of the proposed disturbance observer-based TSMC have been obtained using the PSO algorithm.

### 7.1. Prescribed Finite Time Disturbance Observe using TSMC

Considering the system (47), the prescribed finite time disturbance observer is designed as "(see (48))"
where $k = 4$, $\beta_0 = 7$, $\varepsilon = 10$, $p_0 = 1$, and $q_0 = 7$ are obtained using the PSO algorithm.

The nonsingular prescribed finite time TSMC is designed for the system (47) as "(see (49))"
where $\alpha_1 = 100, \beta_1 = 9, \delta = 5, \mu = 0.0001, p_1 = 3, q_1 = 5, p_2 = 1$, and $q_2 = 3$. $y_d = 0$ is the desired output for the nanobeam which corresponds to minimization of the vibration amplitude. Based on Lemma 2, $p_1/q_1$ must satisfy (5) to avoid the singularity problem in the control law (49) for $x_1 = 0$. Considering $n = 2$ and $j = 1$, (5) is given as

$$\frac{p_1}{q_1} > \frac{1}{2} \tag{50}$$

Simulations of the proposed controller are done in the presence of external disturbance $d = 2\sin(0.1\,\pi t) + 3\sin(0.2\sqrt{t+1})$ and initial conditions $x_1 = 1$ and $x_2 = 5$.

$$K_2 = \frac{\frac{\bar{A}_{xx}}{2}\int_0^1 (\phi')^2 d\bar{x} \cdot \int_0^1 \phi''\phi d\bar{x} - \beta^2 \bar{A}_{xx}\left(\int_0^1 \phi'''\phi' \, d\bar{x} \cdot \int_0^1 \phi''\phi d\bar{x} + \int_0^1 (\phi'')^2 \, d\bar{x} \cdot \int_0^1 \phi''\phi d\bar{x}\right)}{\alpha^2 \bar{I}_A \int_0^1 \phi''\phi \, d\bar{x} - \bar{I}_A \int_0^1 (\phi)^2 \, d\bar{x}}$$
$$- \frac{\frac{\alpha^2 \bar{A}_{xx}}{2}\int_0^1 (\phi')^2 d\bar{x} \cdot \int_0^1 \phi^{(4)}\phi \, d\bar{x} - \alpha^2 \beta^2 \bar{A}_{xx}\left(\int_0^1 \phi'''\phi' \, d\bar{x} \cdot \int_0^1 \phi^{(4)}\phi \, d\bar{x} + \int_0^1 (\phi'')^2 \, d\bar{x} \cdot \int_0^1 \phi^{(4)}\phi \, d\bar{x}\right)}{\alpha^2 \bar{I}_A \int_0^1 \phi''\phi \, d\bar{x} - \bar{I}_A \int_0^1 (\phi)^2 \, d\bar{x}}$$
(45)

$$\begin{aligned}
s &= z - x_2 \\
\dot{z} &= -ks - \beta_0 \, sgn(s) - \varepsilon s^{p_0/q_0} - |-K_1 x_1 - K_2 x_1^3| \, sgn(s) - gu \\
\hat{d} &= -ks - \beta_0 \, sgn(s) - \varepsilon s^{p_0/q_0} - |-K_1 x_1 - K_2 x_1^3| \, sgn(s) + (K_1 x_1 + K_2 x_1^3)
\end{aligned}$$
(48)

$$\begin{aligned}
s_1 &= x_1 - y_d = x_1 \\
s_2 &= \dot{s}_1 + \alpha_1 s_1 + \beta_1 s_1^{p_1/q_1} + s \\
u &= -\frac{1}{g}\left(K_1 x_1 + K_2 x_1^3 + \alpha_1 x_2 + \beta_1 \frac{d}{dt} x_1^{p_1/q_1} + \hat{d} + \delta s_2 + \mu s_2^{p_2/q_2}\right)
\end{aligned}$$
(49)

Fig. 3a shows the convergence of the states of nanobeam to zero in prescribed finite time. Fig. 3b and Fig. 3c depict the trajectory in the phase plane and the designed control input, respectively. Fig. 3d illustrates the deflection of the nanobeam stabilized by the designed control scheme.

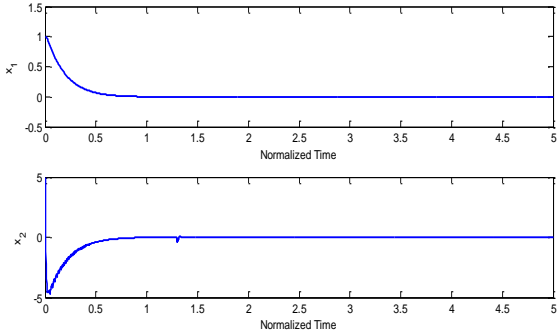

***Fig. 3a.*** *The states of closed-loop nanobeam system for the designed controller (49).*

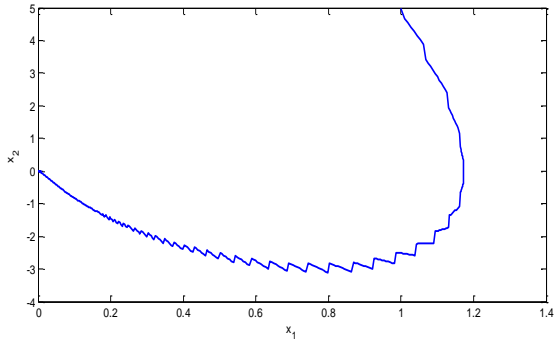

***Fig. 3b.*** *The trajectory in the phase plane for the designed controller (49).*

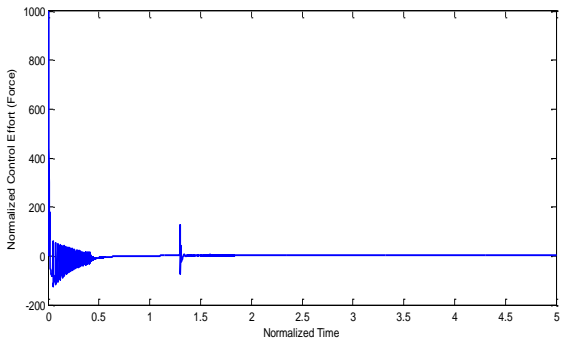

***Fig. 3c.*** *The control input for the designed controller (49).*

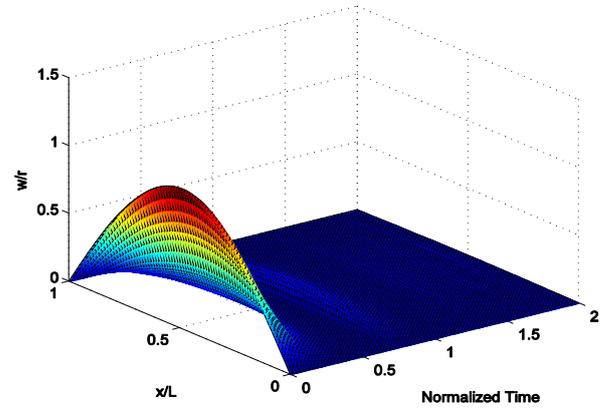

***Fig. 3d.*** *Deflection of the nanobeam stabilized by the designed controller (49).*

### 7.2. Prescribed Finite Time Disturbance Observer using TSMC with Non-symmetric Input Saturation

The prescribed finite time disturbance observer is "(see (51))"
where $k = 5, \beta_0 = 6, \varepsilon = 10, p_0 = 1$, and $q_0 = 7$.

The nonsingular prescribed finite time TSMC with the unknown non-symmetric input saturation is designed as "(see (52))"
where $\alpha_1 = 4.9, \beta_1 = 3, \delta = 3, \mu = 0.01, \tau = 3.7, p_1 = 3, q_1 = 5, p_2 = 1$, and $q_2 = 3$. $y_d = 0$ is the desired output for the nanobeam.

The system is simulated for the unknown non-symmetric input saturation (18) in the presence of external disturbance $d = 0.2\sin(0.1\,\pi t) + 0.3\sin(0.2\sqrt{t+1})$, saturation bounds $u_{min} = -30$, $u_{max} = 10$, and initial conditions $x_1 = 1$ and $x_2 = 5$. Fig. 4a displays the convergence of the states of nanobeam to zero in prescribed finite time. The stabilization time is indirectly related to the values of saturation bounds. Decreasing the bounds, the stabilization time is increased and vice versa. Fig. 4b and Fig. 4c show the closed-loop system trajectory in the phase plane and the designed system control input, respectively. Fig. 4d plots the deflection of nanobeam stabilized by the designed controller. Fig. 5a and Fig. 5b depict the sliding surface $s_2$ for the both cases (i.e. without input saturation and with input saturation), respectively.

$$s = z - x_2$$
$$\dot{z} = -ks - \beta_0 \, sgn(s) - \varepsilon s^{p_0/q_0} - |-K_1 x_1 - K_2 x_1^3| \, sgn(s) + v_r \quad (51)$$
$$\hat{D} = -ks - \beta_0 \, sgn(s) - \varepsilon s^{p_0/q_0} - |-K_1 x_1 - K_2 x_1^3| \, sgn(s) + (K_1 x_1 + K_2 x_1^3)$$

$$s_1 = x_1 - y_d = x_1$$
$$s_2 = \dot{s}_1 + \alpha_1 s_1 + \beta_1 s_1^{p_1/q_1} + s \quad (52)$$
$$v_r = (K_1 x_1 + K_2 x_1^3) - \alpha_1 x_2 - \beta_1 \frac{d}{dt} x_1^{p_1/q_1} - \hat{d} - \delta s_2 - \mu s_2^{p_2/q_2}$$

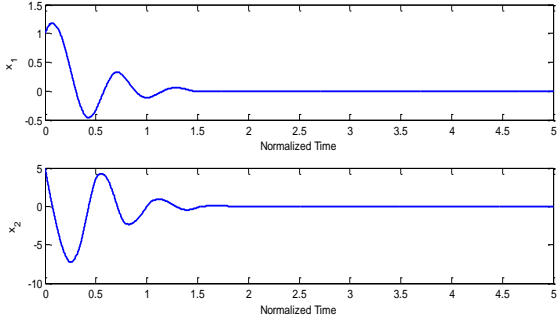

*Fig. 4a. The states of closed-loop nanobeam system for the designed controller (52).*

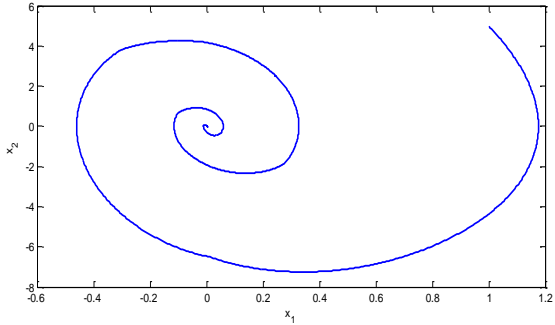

*Fig. 4b. The trajectory in the phase plane for the designed controller (52).*

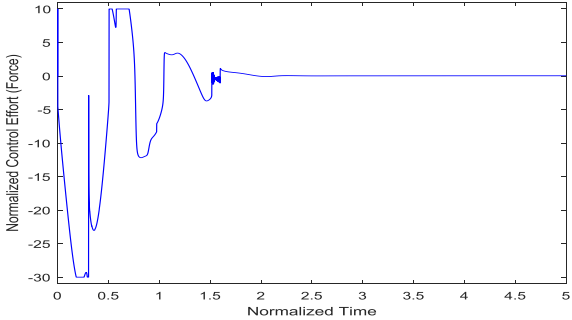

*Fig. 4c. The control input for the designed controller (52).*

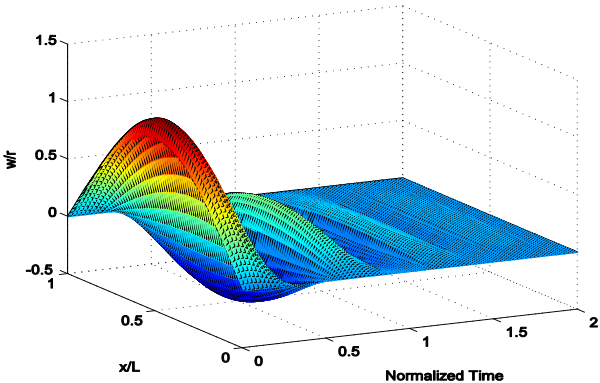

*Fig. 4d. Deflection of the nanobeam stabilized by the designed controller (52).*

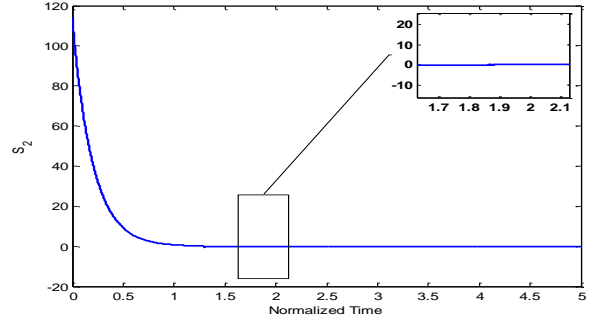

*Fig. 5a. $s_2$ without input saturation.*

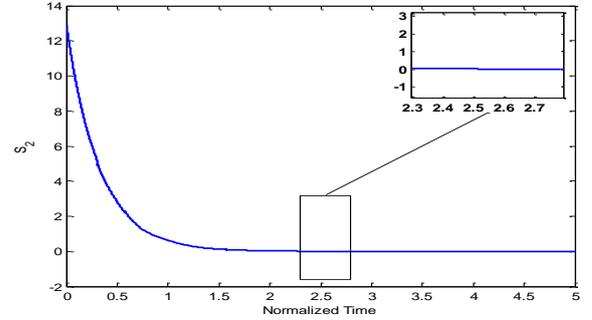

*Fig. 5b. $s_2$ with input saturation.*

### 7.3. Adaptive Nonsingular Prescribed Finite Time Disturbance Observer-based TSMC

The EKF algorithm is utilized to estimate the unknown parameter $K_1$ and the state vector $x$. Therefore, the following augmented state vector and equations are considered as

$$\hat{x} = \begin{bmatrix} \hat{x}_1 \\ \hat{x}_2 \\ \hat{K}_1 \end{bmatrix} \quad (53)$$

$$q = 0.01$$
$$r = 0.01 \quad (54)$$

$$Q = \begin{bmatrix} (0.01)^2 & 0 & 0 \\ 0 & (0.01)^2 & 0 \\ 0 & 0 & 0.01 \end{bmatrix}, \quad R = 0.01 \quad (55)$$

$$[F_k] = \begin{bmatrix} 1 & T_s \\ T_s(-\hat{K}_1 - 3K_2 \hat{x}_{1k}^2) & 1 \end{bmatrix} \quad (56)$$
$$[H_k] = [1 \quad 0]$$

Fig. 6 demonstrates the performance of the adaptive nonsingular prescribe finite time disturbance observer-based TSMC and the EKF algorithm in the presence of external disturbance, system and measurement noises, and input saturation. Initial values of the estimated states and parameters are $\hat{x}_1 = 1$, $\hat{x}_2 = 5$, and $\hat{K}_1 = 20$. Fig. 6a and Fig. 6b display the convergence of the estimated states $\hat{x}_1$ and $\hat{x}_2$ to the real states of the noisy system, respectively. Fig. 6c depicts the convergence of the estimated parameter $\hat{K}_1$ to the actual value of the unknown parameter $K_1$.

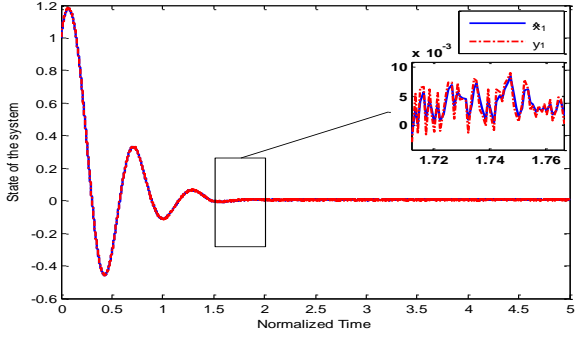

**Fig. 6a.** *The estimated state $\hat{x}_1$ and the real state $y_1$ of the noisy system.*

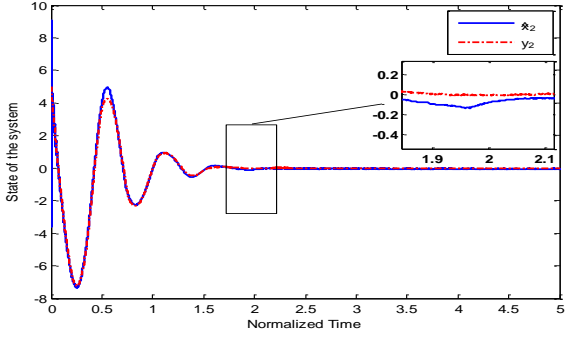

**Fig. 6b.** *The estimated state $\hat{x}_2$ and the state $y_2$ of the noisy system.*

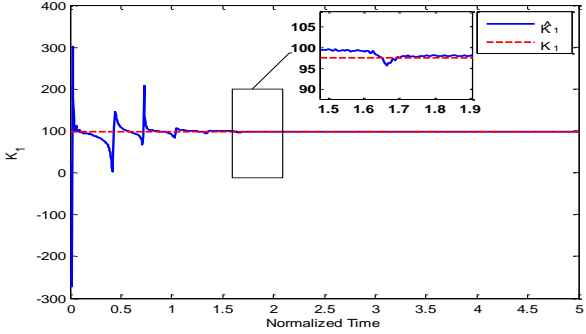

**Fig. 6c.** *The estimated parameter $\hat{K}_1$ and the unknown parameter $K_1 = 97.4$.*

### 7.4. Verification via SMC

In this section, the SMC is employed for the considered problem to verify the performance of the designed control scheme by comparing the numerical results of both controllers.

According to the basic SMC, the sliding surface is obtained for the considered nanobeam (47) as follows

$$s = \left(\frac{d}{dt} + \Upsilon\right)^{n-1} e \xrightarrow{n=2} s = \dot{e} + \Upsilon e \quad (57)$$

where $e$ and $\Upsilon$ represent the tracking error and a design positive parameter, respectively.

Based on the SMC, the following control law is given for the nonlinear system (47) with the uncertain parameter $K_{1\,min} < K_1 < K_{1\,max}$ and $d(\bar{t}) = 0$ as "(see (58-60))"

$$u = u_{eq} + u_c \quad (58)$$

$$\dot{s} = \Upsilon x_2 - \hat{K}_1 x_1 - K_2 x_1^3 - g u_{eq} = 0 \to u_{eq} = \frac{1}{g}(\Upsilon x_2 - \hat{K}_1 x_1 - K_2 x_1^3) \quad (59)$$

where $\eta > 0$ and $K \geq \eta$.

Considering the previous parameters $K_1 = 97.4$, $K_2 = -19.97$, and $g = -1.09$, the range of the uncertain coefficient $K_1$ is assumed as $94.8 < K_1 < 100$. Using the SMC (58), Fig. 7a shows the states of the nanobeam which converge to zero. Figs. 7b and 7c depict the trajectory in the phase plane and the control input, respectively. The deflection of nanobeam is plotted in Fig 7d.

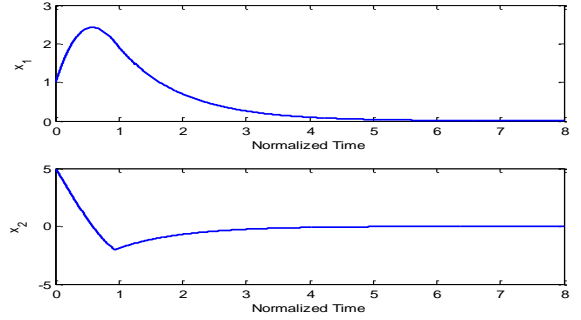

**Fig. 7a.** *The states of closed-loop nanobeam system for the designed controller (52).*

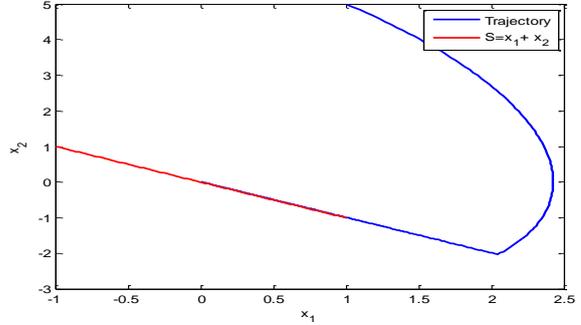

**Fig. 7b.** *The trajectory in the phase plane for the designed controller (52).*

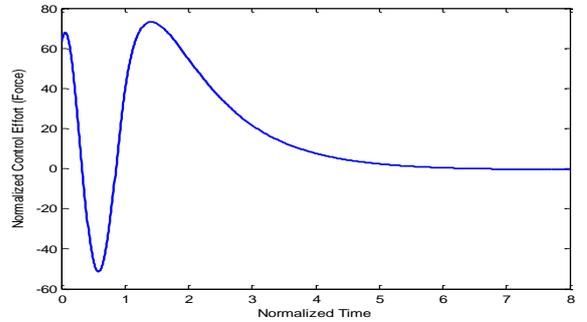

**Fig. 7c.** *The control input for the designed controller (52).*

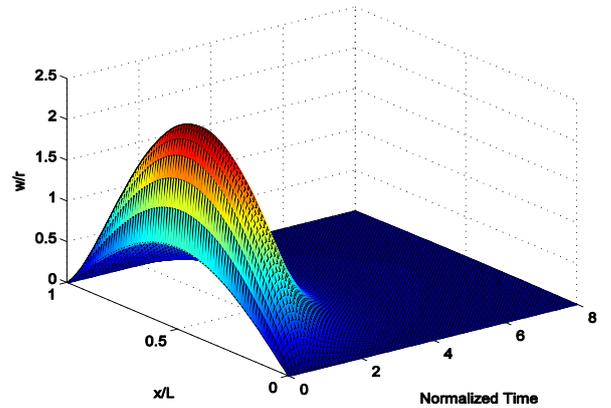

**Fig. 7d.** *Deflection of the nanobeam stabilized by the designed controller (52).*

**Table 2** Comparison of four controllers, i.e. TSMC, TSMC with input saturation, adaptive TSMC with input saturation, and SMC (For the first, second, and forth rows $e_y = y - y_d$, and for the third row $e_x = x_1 - \hat{x}_1$ and $e_y = y - y_d$.

|  | $\|u\|_2$ | $\|u\|_\infty$ | $\|e_y\|_2$ | $\|e_y\|_\infty$ | $\|e_x\|_2$ | $\|e_x\|_\infty$ | $t_s$ |
|---|---|---|---|---|---|---|---|
| TSMC without input saturation | 1316.9 | 999.3451 | 6.6577 | 1.0117 | – | – | 1.9 |
| TSMC with input saturation | 399.9741 | 43.1235 | 17.0777 | 1.1759 | – | – | 2.6 |
| ATSMC with input saturation | 229.5822 | 100 | 17.1464 | 1.1837 | 0.0083 | 0.0083 | – |
| SMC | 894.6961 | 73.1680 | 25.8225 | 2.4259 | – | – | 6.7 |

$$s\dot{s} \leq -\eta|s| \rightarrow s[\Upsilon x_2 - K_1 x_1 - K_2 x_1^3 - g(u_{eq} + u_c)] = s[(\hat{K}_1 - K_1)x_1 - gu_c]$$
$$\leq |s|(\hat{K}_1 - K_{1\min})|x_1| - sgu_c \leq -\eta|s| \rightarrow u_c = \frac{(\hat{K}_1 - K_{1\min})|x_1| + K}{g} \times sgn(s) \quad (60)$$

Table 2 lists the convergence time $t_s$ and the norms of control input and estimation error for all controllers simulated in this study. $\|.\|_2$ stands for the Euclidian norm, and $\|.\|_\infty$ denotes the infinity norm. It is worth noting that considering the input saturation in the control design is necessary for the nanobeam due to the physical constraints. As shown in Table 2, the value of control input with input saturation is less than the cases without input saturation but the convergence time is more. Moreover, it is obvious that the stabilization time and the tracking error of the TSMC are smaller than the SMC.

## 8. Conclusion

In this paper, an adaptive nonsingular prescribed finite time disturbance observer-based TSMC was designed to stabilize the SISO uncertain nonlinear systems in the presence of unmeasurable states, unknown parameters, external disturbance, and non-symmetric input saturation. The EKF algorithm was utilized to estimate the states and unknown parameters of the system. The designed parameters of the proposed disturbance observer-based TSMC were achieved using the PSO algorithm. The suppression of the nonlinear vibration of nonlocal strain gradient nanobeam was investigated using the proposed control scheme. The Galerkin approach was employed to reduce the governing partial nonlinear differential equation of motion to the ordinary nonlinear differential equation with cubic nonlinearity. The formula of upper bound of the convergence time was derived for the considered nanobeam. The numerical simulation results were presented to illustrate the effectiveness and performance of the TSMC and the EKF for stabilization of the nonlinear vibration of nanobeam. It is worth noting that, the main problem of TSMC is the existence of singularity in the control law, where the considered condition for design parameters eliminated this problem. However, as the future work, inspired by [44,45], authors consider the design of a data-based disturbance observer using integral terminal sliding mode control to eliminate the singularity problem in control law and estimate the internal and external disturbances of the system. Moreover, due to the discontinuous function sgn(.) appearing in the disturbance observer and control law, the chattering phenomenon may occur; therefore, a fuzzy logic (FL) is considered to deal by this problem.